\documentclass[aps,prl,twocolumn,epsfig,groupedaddress]{revtex4}

\usepackage{epsfig}
\begin{document}

\title{Distorting the Hump-backed Plateau of Jets with Dense QCD Matter}

\author{Nicolas Borghini$^1$ and Urs Achim Wiedemann$^{1,2}$}
\address{$^1$ Physics Department, Theory Division, CERN, CH-1211 Geneva 23, Switzerland\\
$^2$Physics Department, University of Bielefeld, D-33501 Bielefeld, Germany}

\date{\today}

\begin{abstract}
The hump-backed plateau of the single inclusive distribution of hadrons inside
a jet provides a standard test of the interplay between probabilistic parton 
splitting and quantum coherence in QCD. The medium-induced modification
of this QCD radiation physics is expected to give access to the properties
of the dense medium produced in relativistic heavy ion collisions. Here,
we introduce a formulation of medium-induced parton energy loss, which
treats all leading and subleading parton branchings equally, 
and which -- for showering in the vacuum -- accounts for the observed distribution 
of soft jet fragments. We show that the strong suppression of single inclusive
hadron spectra measured in Au-Au collisions at the Relativistic Heavy Ion
Collider (RHIC) implies a characteristic distortion of the hump-backed plateau; 
we determine, as a function of jet energy, to what extent the
soft jet fragments can be measured above some
momentum cut. Our study further indicates that the approximate
$p_T$-independence of the measured nuclear modification factor does not
exclude a significant $Q^2$-dependence of parton energy loss.
\end{abstract}
\maketitle
 \vskip 0.3cm
%%%%%%%%%%%%%%%%%%%%%%%%%%%%%%%%%%%%%%%%%%%%%%%%%%%%%%%%%%%%%%%%%%%%%%

Over the last five years, experiments at RHIC have established a phenomenon of  
strong high-$p_T$ 
hadron suppression~\cite{Adcox:2004mh,Adams:2005dq}, which 
supports the picture that high-$p_T$ partons produced in the dense matter
of a nuclear collision suffer a significant energy degradation prior to
hadronization in the vacuum~\cite{Jacobs:2004qv}. The microscopic 
dynamics conjectured to underly high-$p_T$ hadron suppression is 
medium-induced gluon radiation~\cite{Baier:2000mf,Kovner:2003zj,Gyulassy:2003mc}, 
i.e., a characteristic medium-induced distortion of the standard
QCD radiation pattern tested extensively by jet measurements in 
high energy $e^+e^-$ and $pp\, (p\bar{p})$ collisions. Modeling this
effect accounts for the measured single-inclusive spectra and leading back-to-back hadron 
correlations~\cite{Jacobs:2004qv,Gyulassy:2003mc,Dainese:2004te,Eskola:2004cr}. 
However, existing models indicate only rough qualitative features of subleading 
jet fragments such as their broadening and softening due to medium-effects. This is so 
mainly, since medium-induced parton splitting is included for 
the leading partons only, but not consistently
for the subleading splitting processes~\cite{Baier:2001yt,Gyulassy:2003mc}, 
and since energy momentum conservation is taken into account globally, but
not ensured locally for each parton splitting~\cite{Gyulassy:2003mc,Eskola:2004cr}. 
Here, we propose a formulation which overcomes these limitations. 

Subleading jet fragments are known to provide many fundamental tests of 
QCD radiation physics. In particular, for soft particle momentum fractions
$x = p/E_{\rm jet}$ inside a quark- or gluon- ($i=q,g$) initiated jet of energy 
and virtuality $Q\sim E_{\rm jet}$, the single inclusive distribution $D_i(x,Q^2)$
is dominated by multiparton destructive interference, and thus tests quantitatively
the understanding of QCD coherence~\cite{Mueller:1982cq,Dokshitzer:1988bq}.
Remarkably, to double and single logarithmic accuracy in 
$\xi \equiv \ln \left[ 1/x \right]$ and $\tau \equiv \ln \left[ Q/\Lambda_{\rm eff} \right]$, 
$\Lambda_{\rm eff} = O(\Lambda_{\rm QCD})$, the effects of this
destructive quantum interference can be accounted for by an angular ordering
prescription of a probabilistic parton cascade with leading order (LO) splitting functions.
The so-called modified leading logarithmic approximation (MLLA) resums these effects and
accounts for the large $\sqrt{\alpha_s}$ next-to-leading corrections of 
$D_i(x,Q^2)$~\cite{Bassetto:1984ik,Mueller:1982cq,Dokshitzer:1988bq}.
The MLLA leads to an evolution equation for the $\nu$-th Mellin moments 
$M_i(\nu,\tau) = \int_0^{\infty} d\xi\, e^{-\nu\xi}\, xD_i(x,Q^2)$
~\cite{Bassetto:1984ik,Dokshitzer:1988bq,Fong:1990nt},
\begin{eqnarray}
\frac{\partial}{\partial \tau}
\left( \begin{array}{c} M_q(\nu,\tau) \\   M_g(\nu,\tau) \end{array} \right)
&=& 
 \left[ \begin{array}{cc}
        \Phi_{qq}\left(\nu + \frac{\partial}{\partial \tau} \right) 
        &  \Phi_{qg}\left(\nu + \frac{\partial}{\partial \tau} \right) \\
        \Phi_{gq}\left(\nu + \frac{\partial}{\partial \tau} \right)  & 
        \Phi_{gg}\left(\nu + \frac{\partial}{\partial \tau} \right)
 \end{array} \right]
        \nonumber \\
        && \qquad \times
         \frac{\alpha_s(\tau)}{2\, \pi}\, 
 \left( \begin{array}{c} M_q(\nu,\tau) \\   M_g(\nu,\tau) \end{array} \right)\, .
 \label{eq1}
\end{eqnarray}
Here, $\Phi_{ij}$ denote 
combinations of particular moments of leading order splitting functions, for example
\begin{eqnarray}
        \Phi_{qq}\left(\nu\right) &=&
        2\int_0^1 dz\, P_{qq}(z)\, \left( z^\nu - 1 \right)\, .
%       \nonumber \\
%        \Phi_{qg}\left(\nu\right) &=&
%       2\int_0^1 dz\, P_{qq}(z)\, \left( 1- z\right)^\nu \, ,
%       \nonumber \\
%         \Phi_{gg}\left(\nu\right) &=&
%       \int_0^1 dz\, \lbrace P_{gg}(z)\, \left[z^\nu + (1-z)^\nu - 1 \right] - P_{qg}(z) \rbrace \, ,
%         \nonumber \\
%       \Phi_{gq}\left(\nu\right) &=&
%       2\int_0^1 dz\, P_{qg}(z)\, \left[ z^\nu + (1- z)^\nu \right] \, .
        \label{eq2}
\end{eqnarray}
The shift $\left( \nu + \frac{\partial}{\partial \tau} \right)$ in (\ref{eq1}) 
accounts for angular ordering. For a parton fragmentation which starts at
high initial scale $\tau$ and ends at some hadronic scale $\tau_0$, the 
solution of (\ref{eq1}) has to fulfill the initial conditions $M(x,\tau_0) = \delta (1-x)$
and $\frac{\partial}{\partial \tau} M(x,\tau=\tau_0) = 0$, since the parton must not
evolve if produced at the hadronic scale. 

\begin{figure}[h]\epsfxsize=8.7cm
%\centerline{\epsfbox{ktmorkap.eps}}
%\centerline{\epsfbox{ktmorom.eps}}
\centerline{\epsfbox{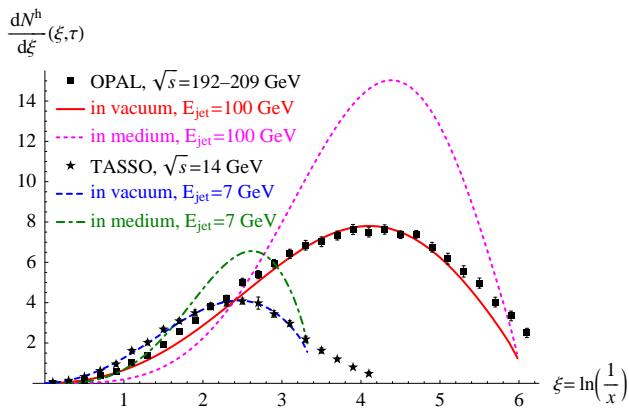}}
%\vspace{0.5cm}
\caption{The single inclusive hadron distribution as a function of 
$\xi = \ln \left[E_{\rm jet}/p\right]$. Data taken from $e^+e^-$ collision
experiments TASSO~\cite{Braunschweig:1990yd} and OPAL~\cite{Abbiendi:2002mj},
$E_{\rm jet} = \sqrt{s}/2$. Lines through data obtained from the
MLLA result (\ref{eq3}). Dashed and dash-dotted curves labeled 
"in medium" are calculated with a medium-modification $f_{\rm med} = 0.8$
of the LO splitting functions.
}\label{fig1}
\end{figure}

The lowest Mellin moments $\nu \sim 0$ determine the main characteristics of $D_i(x,Q^2)$.
For an approximate solution of (\ref{eq1}), one can thus expand the matrix in (\ref{eq1}) to 
next-to-leading order in $\left( \nu + \frac{\partial}{\partial \tau} \right)$ and diagonalize it.
Its eigenvalue with leading $1/\left( \nu + \frac{\partial}{\partial \tau} \right)$-term
yields a differential equation of the confluent hypergeometric type~\cite{Dokshitzer:1988bq}. 
This leads to an analytic expression for $D(x,Q^2)$, whose shape does not distinguish 
between quark and gluon parents, since the multiplicity is dominated in both cases
by gluon branching. For the hadronic multiplicity distribution
$dN^h/d\xi$, one assumes that at the scale $\tau_0$, a parton
is mapped locally onto a hadron with proportionality factor $K^h\sim O(1)$
("local parton hadron duality", LPHD)
\begin{equation}
       \frac{dN^h}{d\xi} = K^h\, D\left(x,\tau=\ln \left[ \frac{Q=E}{\Lambda_{\rm eff}}\right] \right)\, .
%       dN^h/d\xi = K^h\, D\left(x,\tau=\ln \left[(Q=E)/\Lambda_{\rm eff} \right] \right)\, .
        \label{eq3}
\end{equation}
Comparisons of (\ref{eq3}) to data have been performed 
repeatedly~\cite{Fong:1990nt,Dokshitzer:1992jv,Braunschweig:1990yd,Abbiendi:2002mj,Acosta:2002gg} over a 
logarithmically wide kinematic regime $7 < E_{\rm jet} < 150$ GeV in both
$e^+e^-$ and $pp/p\bar{p}$ collisions. To illustrate the degree of agreement,
we reproduce in Fig.~\ref{fig1} two sets of data~\cite{Braunschweig:1990yd,Abbiendi:2002mj} 
together with the curves obtained from (\ref{eq3}). The parameters $K^h$ and $\Lambda_{\rm eff}$ 
entering (\ref{eq3}) were chosen as in Refs.~\cite{Braunschweig:1990yd,Abbiendi:2002mj}, 
$\Lambda_{\rm eff} = 254$ MeV, $K^h= 1.15$ for $E_{\rm jet} = 100$ GeV, 
$K^h= 1.46$ for $E_{\rm jet} = 7$ GeV. Following  Ref.~\cite{Abbiendi:2002mj}, we 
use $N_f = 3$. From Fig.~\ref{fig1}, we conclude that Eq.(\ref{eq3}) accounts 
reasonably well for the jet multiplicity distribution in the kinematic range accessible
in heavy ion collisions at RHIC ($E_{\rm jet} \sim 10$ GeV) and at the LHC 
($E_{\rm jet} \sim 100$ GeV). 
Corrections not included in (\ref{eq3}) are of relative order $1/\tau$, which at face value
corresponds to a 30\% (15\%) uncertainty at typical RHIC (LHC) jet energies. Also,
the MLLA resums large $\xi$, $\tau \sim \xi$, but is expected to be less accurate for hard 
jet fragments, where other improvements are currently sought for~\cite{Albino:2005gg}. 
Thus, the agreement of (\ref{eq3}) to data for the
entire $\xi$-range is surprisingly good. At least from a pragmatic point of view,
(\ref{eq3}) can serve as a baseline on top of which one can search for 
medium effects. 
%
%%%%%%%%%%%%%%%%%%%%%%%%%%%%%%%%%%%%%%%%%%%%%%%%%%%%%%%%%%%%%%%%%%%%
%%%%%%%%%%%%%%%%%%%%%%%%%%%%%%%%%%%%%%%%%%%%%%%%%%%%%%%%%%%%%%%%%%%

The multiplicity distribution $dN^h/d\xi$ is dominated by soft
gluon bremsstrahlung, $dI^{\rm vac} \simeq C_R \frac{\alpha_s(k_T^2)}{\pi}
\frac{dk_T^2}{k_T^2}\, \frac{d\omega}{\omega}$, $\omega = z\, E_{\rm jet}$, which
is described by the singular parts $\sim \frac{1}{z}$, 
$\sim \frac{1}{(1-z)}$ of the QCD splitting functions entering (\ref{eq2}).
They determine the leading $\frac{1}{\nu}$-terms of the evolution matrix
in (\ref{eq1}). Remarkably, calculations of the additional 
medium-induced radiation indicate that
$\omega \frac{dI^{\rm med}}{d\omega}$ is  $\sim \frac{1}{\sqrt{\omega}}$ if 
the medium is modeled by soft multiple momentum 
transfers~\cite{Baier:1996sk,Salgado:2003gb}, and 
% $\omega \frac{dI^{\rm med}}{d\omega} 
$\sim \frac{1}{\omega}$
if the medium is modeled by a single hard momentum 
transfer~\cite{Gyulassy:2003mc,Salgado:2003gb}. Thus,
parametrically, the additional medium-dependent contributions to the
gluon bremsstrahlung are more singular than $dI^{\rm vac}$ for small $\omega$
and may thus be expected to dominate the multiplicity distribution (\ref{eq3}).
However, destructive interference due to finite in-medium path length is known
to regulate the soft $\omega$-divergence~\cite{Salgado:2003gb}. For
the relevant range of soft $\omega$, this may be modeled as
$\omega \frac{dI^{\rm med}}{d\omega} \sim f_{\rm med} = {\rm const}$.
A medium-induced gluon bremsstrahlung spectrum, consistent with this ansatz,
was also found in~\cite{Guo:2000nz}. This suggests that medium effects 
enter (\ref{eq3}) by enhancing the singular parts of all LO splitting functions
$P_{gg}$, $P_{qg}$, $P_{qq}$ by the same factor $\left(1+f_{\rm med} \right)$, 
such that for example
\begin{equation}
        P_{qq}(z) = C_F\, \left(\frac{2\, \left(1+f_{\rm med}\right)}{(1-z)_+}
                                                - (1+z) \right)\, .
        \label{eq4}
\end{equation}
We do not modify the non-singular subleading terms.
On general grounds, one expects that medium-induced rescattering is a nuclear
enhanced higher-twist contribution ($f_{\rm med} \sim \frac{L}{Q^2}$)~\cite{Luo:1993ui}.
This means that it is subleading in an expansion in $Q^2$, while being enhanced 
compared to other higher twist contributions
by a factor proportional to the geometrical extension $\sim L$ of the target. 
A $1/Q^2$-dependence of $f_{\rm med}$ is also
suggested by the following heuristic argument. A hard parton of
virtuality $Q$ has a lifetime $\sim 1/Q$ in its own rest frame, and thus a lifetime
(in-medium path length) $t = \frac{1}{Q} \frac{E}{Q}$ before it branches in the rest
frame of the dense matter through which it propagates. Medium effects on a parton 
in between two branching processes should grow proportional to (some power of) 
the in-medium path length and thus $\propto 1/Q^2$ or higher powers thereof.

In contrast, jet quenching models~\cite{Jacobs:2004qv,Gyulassy:2003mc,Dainese:2004te,Eskola:2004cr} reproduce inclusive hadron spectra in Au-Au collisions at RHIC by supplementing
the standard QCD LO factorized formalism with the probability $P(\Delta E)$ that the
produced partons radiate an energy $\Delta E$ due to medium effects prior to 
hadronization in the vacuum~\cite{Baier:2001yt}
\begin{eqnarray}
  P(\Delta E) &=& \sum_{n=0}^\infty \frac{1}{n!}
  \left[ \prod_{i=1}^n \int d\omega_i \frac{dI^{\rm med}(\omega_i)}{d\omega}
    \right]
    \delta\left(\Delta E - \sum_{i=1}^n \omega_i\right)
    \nonumber \\
    && \times \exp\left[ - \int_0^\infty \hspace{-0.2cm}
      d\omega \frac{dI^{\rm med}}{d\omega}\right]\, .
   \label{eq5}
\end{eqnarray}
This formula is based on a probabilistic iteration of medium-modified parton splittings, 
but does not keep track of virtuality or angular ordering. The $k_T$-integrated
medium-induced contribution $dI^{\rm med}$ is treated on an equal footing with
LO vacuum splitting functions.
In this sense, the medium-modified fragmentation function
$  D_{h/q}^{(\rm med)}(x,Q^2) = \int_0^1 d\epsilon\, E\, P(\Delta E)\,
  \frac{1}{1-\epsilon}\,  D_{h/q}(\frac{x}{1-\epsilon},Q^2)$, $\epsilon = \Delta E/E$,
entering jet quenching 
models~\cite{Jacobs:2004qv,Gyulassy:2003mc,Dainese:2004te,Eskola:2004cr}, 
amounts to a medium-induced $Q^2$-independent
modification of parton fragmentation. 

The single inclusive distribution $D(x,Q^2)$, supplemented by LPHD, is a
fragmentation function. Single inclusive hadron spectra, whose parent partons
show a power law spectrum $\propto 1/p_T^{n(p_T)}$, test $D(x,Q^2)$
in the range, in which $x^{n(p_T)-2}\, D(x,Q^2)$ has significant support. 
However, for large $x$, the accuracy of the MLLA result 
for $D(x,Q^2)$ becomes questionable~\cite{Albino:2005gg,Fong:1990nt}. To
understand to what extent the MLLA result may still be used from a practical point
of view, we have compared it to the KKP parametrization~\cite{Kniehl:2000fe} 
of fragmentation functions. In the range of $Q^2$ and $x$ relevant for single 
inclusive spectra ($0.4 < x < 0.9$), we observe that the KKP and
MLLA fragmentation functions both drop by 2 orders of magnitude. They do 
show somewhat different shapes but -- after adjusting the overall normalization -- 
they differ for all $x$-values by maximally $\sim 30$\% or significantly 
less (data not shown). For the nuclear modification factor $R_{AA}$, which is 
the ratio of modified and unmodified single inclusive hadron spectra, and which
does not depend on the overall normalization of $D(x,Q^2)$, this is a relatively
small uncertainty, if one aims at characterizing a factor 5 suppression. We thus 
conclude that the MLLA fragmentation function obtained from (\ref{eq1}) can be 
used to calculate $R_{AA}$.

%%%%%%%%%%%%%%%%%%%%%%%%%%%%%%%%%%%%%%%%%%%%%%%%%%%%%%%%%%%%%%%%%%%%
\begin{figure}[h]\epsfxsize=8.5cm
\centerline{\epsfbox{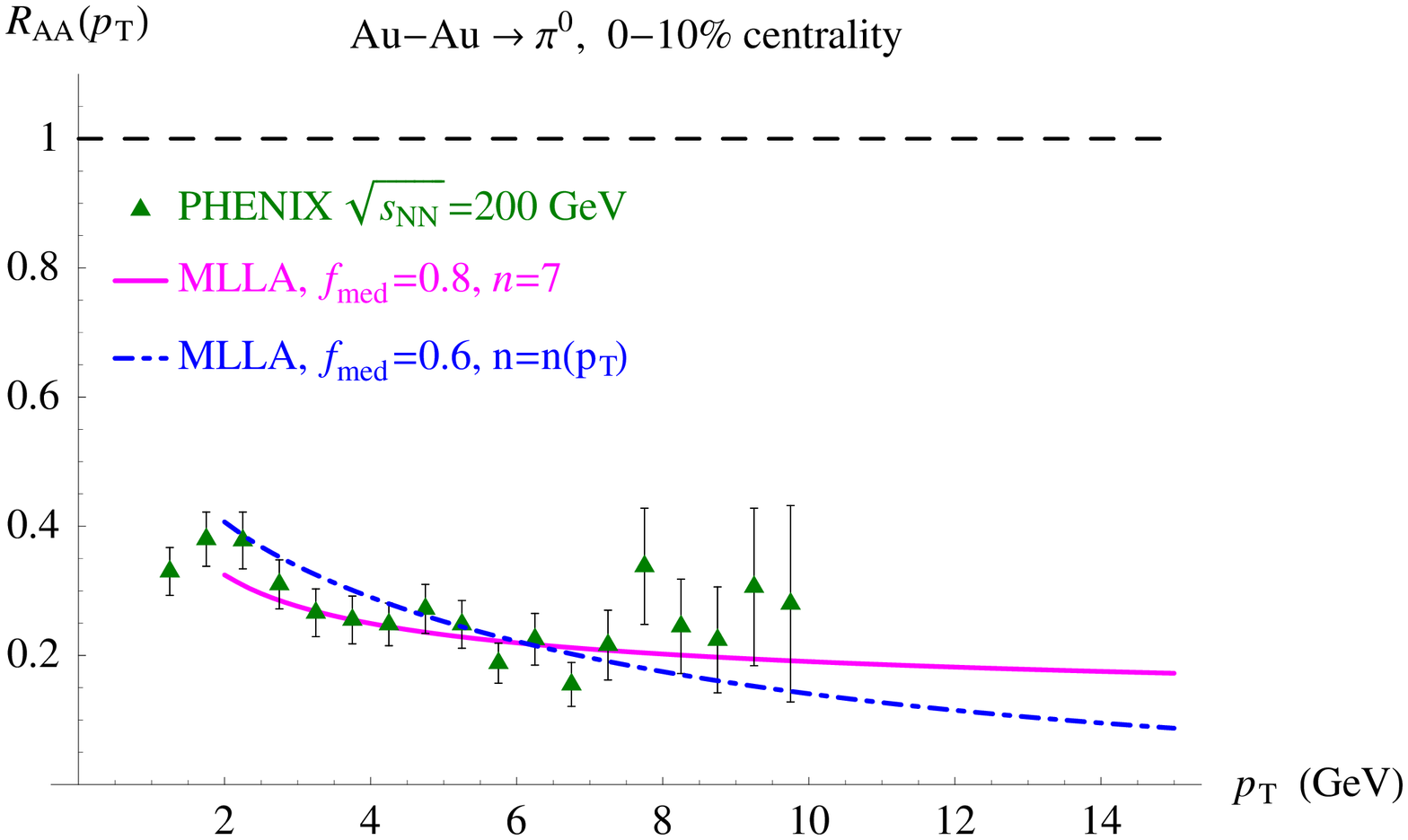}}
%\vspace{0.5cm}
\caption{The $p_T$-dependence of the nuclear modification factor $R_{AA}$,
calculated for a medium-enhanced 
parton splitting with $f_{\rm med}$. Data taken from~\cite{Adler:2003ii}.
}\label{fig2}
\end{figure}
%%%%%%%%%%%%%%%%%%%%%%%%%%%%%%%%%%%%%%%%%%%%%%%%%%%%%%%%%%%%%%%%%%%
To determine $R_{AA}$, we have parametrized the partonic $p_T$-spectrum at 
RHIC energies $\sqrt{s_{NN}} = 200$ GeV by a power law 
$1/p_T^{n(p_T)}$, $n(p_T) = 7 + 0.003 p_T^2/{\rm GeV}^2$, which
accounts for kinematic boundary effects at $p_T \sim O(\sqrt{s_{NN}})$. Single 
inclusive hadron spectra and $R_{AA}$ are calculated by convoluting this
spectrum with $D(x,Q^2)$. Medium effects are included through the factor
$f_{\rm med}$ in the singular parts of all LO parton splitting functions, see Eq.(\ref{eq4}). 
As seen in Fig.~\ref{fig2}, the choice $f_{\rm med} \simeq 0.6 \div 0.8$ 
reproduces the size of the suppression of $R_{AA} \sim 0.2 $ in central Au-Au
collisions at RHIC~\cite{Adler:2003ii}. Jet quenching models based on (\ref{eq5}) 
yield a slightly increasing $p_T$-dependence of $R_{AA}(p_T)$ for a power law
$n(p_T) = n$, and a rather flat dependence if  trigger bias effects due to the 
$p_T$-dependence of $n(p_T)$ are included~\cite{Dainese:2004te,Eskola:2004cr}.
In contrast, the MLLA result for $R_{AA}(p_T)$ decreases with increasing $p_T$, and this
tendency is even more pronounced for a realistic shape of the underlying  partonic
spectrum, see Fig.~\ref{fig2}. The reason is that in the MLLA, parent partons of higher
$p_T$ have higher initial virtuality $Q \sim p_T$, and undergo more
medium-induced splittings; this results in a smaller value of $R_{AA}$. A
proper treatment  of nuclear geometry may affect quantitative aspects of Fig.~\ref{fig2},
but is unlikely to change this qualitative observation. Hence, the
observed flat $p_T$-dependence of $R_{AA}(p_T)$, one may require (in accordance
with the arguments given above) a non-vanishing
$Q^2$-dependence of $f_{\rm med}$, which would reduce medium-effects on 
high-$p_T$ ($p_T \sim Q$) partons. 

Motivated by this observation, we have attempted to solve Eq.(\ref{eq1}) for a non-trivial 
$Q^2$-dependence of the medium-enhancement $f_{\rm med}$.
We did not find an analytical solution. However, 
in the absence of medium effects, the analytical solution (\ref{eq3}) is
reproduced by Monte Carlo (MC) parton showers
based on angular ordering~\cite{Fong:1990nt}. This remains true for
non-vanishing $f_{\rm med}$. The present study
can serve to check future MC showers
implementing (\ref{eq4}), and it can be extended in MC studies to include
a non-trivial $Q^2$-dependence of $f_{\rm med}$. We plan such a MC 
study, mainly to establish to what extent the approximate $p_T$-independence
of $R_{AA}$ up to $p_T\sim 10$ GeV allows for a significant $Q^2$-dependence
of parton energy loss. The question of whether and on what scale these
effects are $1/Q^2$-suppressed is of obvious importance for heavy ion collisions 
at the LHC, where medium-modified 
parton fragmentation can be tested in a logarithmically wide $Q^2$-range.

What is the distortion of the longitudinal jet multiplicity distribution (\ref{eq3}),
consistent with the observed factor $\sim 5$ suppression of $R_{AA}$? 
In contrast to calculations based on (\ref{eq5}), the medium-enhanced parton 
splitting introduced via MLLA conserves energy-momentum exactly 
at each branching, it treats all secondary branchings of softer 
gluons equally, and it continues all branchings down to the same
hadronic scale. 
This makes it a qualitatively improved tool for the calculation of longitudinal
multiplicity distributions, since it matters obviously for $dN/d\xi$ whether
one gluon is radiated into the bin $\xi = \ln \left[ E_{\rm jet}/p_{\rm gluon}\right]$,
or -- after further splitting $g \to g(z)\, g(1-z)$ -- two gluons with momentum
fractions $z$ and $(1-z)$ into bins $\xi + \ln \left[ 1/(1-z)\right]$, 
$\xi +  \ln \left[ 1/z\right]$, respectively. 
We have calculated $dN/d\xi$ for a medium-enhanced
parton splitting $f_{\rm med} = 0.8$ consistent with $R_{AA} = 0.2$.
Results for jet energies relevant at RHIC and at the LHC are shown
in Fig.~\ref{fig1}.  In general, the multiplicity at large momentum fractions
(small $\xi$) is reduced and the corresponding energy
is redistributed into the soft part of the distribution. The maximum of the
multiplicity distribution also shifts to a softer value, but this shift
is subleading in $\sqrt{\alpha_s}$,  $\xi_{\rm max}/\tau = \frac{1}{2} 
+ a_{\rm med} \sqrt{\frac{\alpha_s(\tau)}{32\, N_c\, \pi}}$, where
$a_{\rm med} = \frac{1}{3}\left(11 + 12 f_{\rm med}\right)\, N_c +
\frac{2}{3} \frac{N_f}{N_c^2}$.
%
%%%%%%%%%%%%%%%%%%%%%%%%%%%%%%%%%%%%%%%%%%%%%%%%%%%%%%%%%%%%%%%%%%%%
\begin{figure}[t!]\epsfxsize=8.7cm
\centerline{\epsfbox{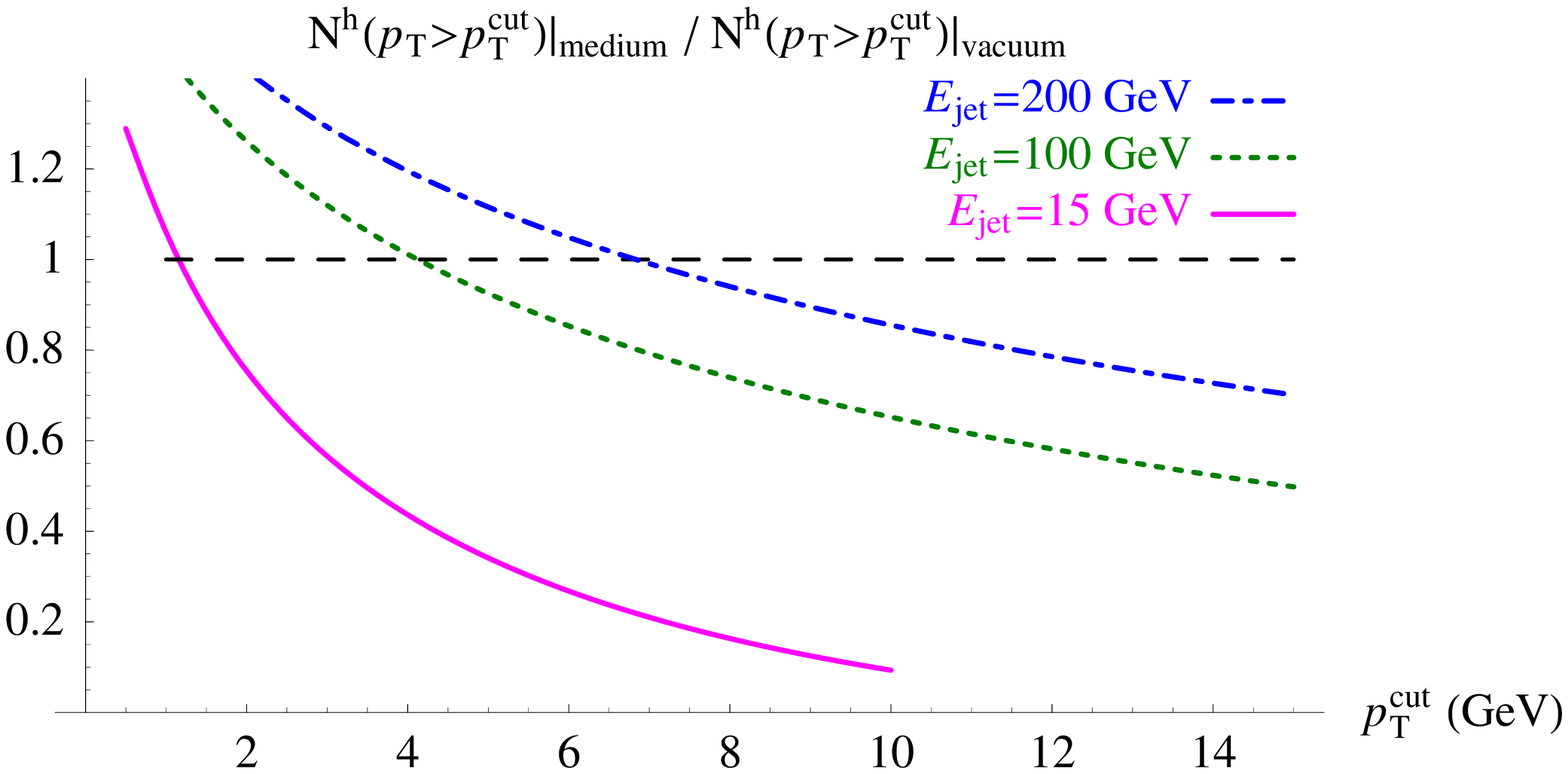}}
%\vspace{0.5cm}
\caption{The change of the total hadronic multiplicity inside jets of different
energy $E_{\rm jet}$, as a function of the soft background cut $p_T^{\rm cut}$ 
above which this multiplicity is measured. Medium-effects are modeled by 
$f_{\rm med} = 0.8$.
}\label{fig3}
\end{figure}
%%%%%%%%%%%%%%%%%%%%%%%%%%%%%%%%%%%%%%%%%%%%%%%%%%%%%%%%%%%%%%%%%%%
%

Many experimental characterizations of the medium-modified internal jet structure in 
heavy ion collisions at RHIC and at the LHC require a soft momentum 
cut $p_T^{\rm cut}$ to control effects of the high multiplicity background. 
Can one observe the increase in soft multiplicity shown in Fig.~\ref{fig1},
if such a soft background cut $p_T^{\rm cut}$ is applied? To address this
issue, we have calculated from (\ref{eq3}) the total hadronic multiplicity 
$N^h(p_T>p_T^{\rm cut})$  above $p_T^{\rm cut}$. As seen in Fig.~\ref{fig3},
the medium-enhanced component of soft multiplicity lies below a critical transverse
momentum cut $p_{T,{\rm crit}}^{\rm cut}$ which increases significantly with $E_{\rm jet}$.
For a typical hard jet at RHIC ($E_{\rm jet} =15$ GeV), the additional soft jet multiplicity
lies buried in the soft background, $p_{T,{\rm crit}}^{\rm cut} \simeq 1.5$ GeV.
For $E_{\rm jet} =100$ GeV, accessible at the LHC, 
$p_{T,{\rm crit}}^{\rm cut} \simeq 4$ GeV lies well above a cut which 
depletes the background multiplicity by a factor 10. For $E_{\rm jet} = 200$ GeV, we find
$p_{T,{\rm crit}}^{\rm cut}\simeq 7$ GeV.
The associated total jet multiplicity $N^h(p_T>p_{T,{\rm crit}}^{\rm cut})$ for
these jet energies rises with $E_{\rm jet}$ from $\simeq 4$, to $\simeq 7$.
Fig.~\ref{fig3} indicates a qualitative advantage in extending jet measurements 
in an LHC heavy ion run near design luminosity to significantly above 
$E_{\rm jet} \simeq 100$ GeV, where a sizeable
kinematic range $2\div 3\, {\rm GeV} < p_T < p_{T,{\rm crit}}^{\rm cut}\simeq 7$ GeV
becomes accessible. This may allow a detailed characterization of the enhanced
medium-induced radiation above the soft background. 

In general, the formulation of parton energy loss within the MLLA formalism
allows one to address several fundamental questions, that remain untouched
by recent model studies of jet quenching. This concerns in particular the important
issue of the $Q^2$-dependence of parton energy loss discussed above. 
Moreover, the use of a probabilistic formulation based
on angular ordering can also be viewed as a test of the unproven assumption
that the medium-induced destructive interference of multi-parton emission can
indeed be accounted for by angular ordering, in close similarity to gluon radiation
in the vacuum. We finally note that the formalism introduced here is not limited
to a discussion of the hump-backed plateau: e.g. it can be applied to the calculation
of two-particle correlations within jets, which have been studied in the absence of 
medium effects~\cite{Bassetto:1984ik,Fong:1990nt}. It may also apply to transverse 
jet broadening~\cite{Bassetto:1984ik}, which for a $Q^2$-dependent $f_{\rm med}$ 
may be significantly reduced since large angle emission would remain essentially
unmodified by the medium. We plan to 
address these open questions in future work.

We thank S. Catani, G. Sterman and B. Webber for helpful discussions.
%
%%%%%%%%%%%%%%%%%%%%%%%%%%%%%%%%%%%%%%%%%%%%%%%%%%%%%%%%%%%%%%%%%%%%%%%


\begin{thebibliography}{9}
%\vspace*{-1.0cm}
%\begin{references}
%
%\cite{Adcox:2004mh}
\bibitem{Adcox:2004mh}
  K.~Adcox {\it et al.}  [PHENIX Collaboration],
  %``Formation of dense partonic matter in relativistic nucleus nucleus
  %collisions at RHIC: Experimental evaluation by the PHENIX  collaboration,''
  arXiv:nucl-ex/0410003.
  %%CITATION = NUCL-EX 0410003;%%
%
%\cite{Adams:2005dq}
\bibitem{Adams:2005dq}
  J.~Adams {\it et al.}  [STAR Collaboration],
  %``Experimental and theoretical challenges in the search for the quark gluon
  %plasma: The STAR collaboration's critical assessment of the evidence from
  %RHIC collisions,''
  arXiv:nucl-ex/0501009.
  %%CITATION = NUCL-EX 0501009;%%
 %
 %\cite{Jacobs:2004qv}
\bibitem{Jacobs:2004qv}
  P.~Jacobs and X.~N.~Wang,
  %``Matter in extremis: Ultrarelativistic nuclear collisions at RHIC,''
  Prog.\ Part.\ Nucl.\ Phys.\  {\bf 54}, 443 (2005).
%  [arXiv:hep-ph/0405125].
  %%CITATION = HEP-PH 0405125;%%
%
%\cite{Baier:2000mf}
\bibitem{Baier:2000mf}
  R.~Baier, D.~Schiff and B.~G.~Zakharov,
  %``Energy loss in perturbative QCD,''
  Ann.\ Rev.\ Nucl.\ Part.\ Sci.\  {\bf 50}, 37 (2000).
%  [arXiv:hep-ph/0002198].
  %%CITATION = HEP-PH 0002198;%%
%
%\cite{Kovner:2003zj}
\bibitem{Kovner:2003zj}
  A.~Kovner and U.~A.~Wiedemann,
  in ``Quark Gluon Plasma 3'', (editors: R.C. Hwa and X.N. Wang,
   World Scientific, Singapore), p.192-248,
  %``Gluon radiation and parton energy loss,''
  arXiv:hep-ph/0304151.
  %%CITATION = HEP-PH 0304151;%%
%
%\cite{Gyulassy:2003mc}
\bibitem{Gyulassy:2003mc}
  M.~Gyulassy, I.~Vitev, X.~N.~Wang and B.~W.~Zhang,
  in ``Quark Gluon Plasma 3'', (editors: R.C. Hwa and X.N. Wang,
   World Scientific, Singapore), p. 123-191,
  %``Jet quenchin %\ g and radiative energy loss in dense nuclear matter,''
  arXiv:nucl-th/0302077.
  %%CITATION = NUCL-TH 0302077;%%
%
%\cite{Dainese:2004te}
\bibitem{Dainese:2004te}
A.~Dainese, C.~Loizides and G.~Paic,
%``Leading-particle suppression in high energy nucleus nucleus collisions,''
Eur.\ Phys.\ J.\ C {\bf 38} (2005) 461.
%[arXiv:hep-ph/0406201].
%%CITATION = HEP-PH 0406201;%%
%
%\cite{Eskola:2004cr}
\bibitem{Eskola:2004cr}
K.~J.~Eskola, H.~Honkanen, C.~A.~Salgado and U.~A.~Wiedemann,
%``The fragility of high-p(T) hadron spectra as a hard probe,''
Nucl.\ Phys.\ A {\bf 747} (2005) 511.
%[arXiv:hep-ph/0406319].
%%CITATION = HEP-PH 0406319;%%
%
%\cite{Baier:2001yt}
\bibitem{Baier:2001yt}
R.~Baier, Yu.~L.~Dokshitzer, A.~H.~Mueller and D.~Schiff,
%``Quenching of hadron spectra in media,''
JHEP {\bf 0109} (2001) 033.
%[arXiv:hep-ph/0106347].
%%CITATION = HEP-PH 0106347;%%
%
%\cite{Mueller:1982cq}
\bibitem{Mueller:1982cq}
  A.~H.~Mueller,
  %``Multiplicity And Hadron Distributions In QCD Jets: Nonleading Terms,''
  Nucl.\ Phys.\ B {\bf 213} (1983) 85.
  %%CITATION = NUPHA,B213,85;%%
%
%\cite{Dokshitzer:1988bq}
\bibitem{Dokshitzer:1988bq}
  Yu.~L.~Dokshitzer, V.~A.~Khoze and S.~I.~Troian,
  %``Coherence And Physics Of QCD Jets,''
  Adv.\ Ser.\ Direct.\ High Energy Phys.\  {\bf 5} (1988) 241.
  %%CITATION = 00319,5,241;%%
%
%\cite{Bassetto:1984ik}
\bibitem{Bassetto:1984ik}
  A.~Bassetto, M.~Ciafaloni and G.~Marchesini,
  %``Jet Structure And Infrared Sensitive Quantities In Perturbative QCD,''
  Phys.\ Rept.\  {\bf 100} (1983) 201.
  %%CITATION = PRPLC,100,201;%%
%
%\cite{Fong:1990nt}
\bibitem{Fong:1990nt}
  C.~P.~Fong and B.~R.~Webber,
  %``One And Two Particle Distributions At Small X In QCD Jets,''
  Nucl.\ Phys.\ B {\bf 355} (1991) 54.
  %%CITATION = NUPHA,B355,54;%%
 %
  %\cite{Dokshitzer:1992jv}
\bibitem{Dokshitzer:1992jv}
  Yu.~L.~Dokshitzer, V.~A.~Khoze and S.~I.~Troian,
  %``Phenomenology of the particle spectra in QCD jets in a modified leading
  %logarithmic approximation,''
  Z.\ Phys.\ C {\bf 55} (1992) 107.
  %%CITATION = ZEPYA,C55,107;%%
  %
\bibitem{Braunschweig:1990yd}
  W.~Braunschweig {\it et al.}  [TASSO Collaboration],
  Z.\ Phys.\ C {\bf 47} (1990) 187.
  %%CITATION = ZEPYA,C47,187;%%
  %
\bibitem{Abbiendi:2002mj}
  G.~Abbiendi {\it et al.}  [OPAL Collaboration],
  Eur.\ Phys.\ J.\ C {\bf 27} (2003) 467.
%  [arXiv:hep-ex/0209048].
  %%CITATION = HEP-EX 0209048;%%
%
%\cite{Acosta:2002gg}
\bibitem{Acosta:2002gg}
  D.~Acosta {\it et al.}  [CDF Collaboration],
  %``Momentum distribution of charged particles in jets in dijet events in p
  %anti-p collisions at s**(1/2) = 1.8-TeV and comparisons to perturbative  QCD
  %predictions,''
  Phys.\ Rev.\ D {\bf 68}, 012003 (2003).
  %%CITATION = PHRVA,D68,012003;%%
%
%\cite{Albino:2005gg}
\bibitem{Albino:2005gg}
  S.~Albino, B.~A.~Kniehl, G.~Kramer and W.~Ochs,
  %``Generalizing the DGLAP evolution of fragmentation functions to the smallest
  %x values,''
  arXiv:hep-ph/0503170.
  %%CITATION = HEP-PH 0503170;%%
%
%\cite{Baier:1996sk}
\bibitem{Baier:1996sk}
R.~Baier, Yu.~L.~Dokshitzer, A.~H.~Mueller, S.~Peign\'e and D.~Schiff,
%``Radiative energy loss and p(T)-broadening of high energy
% partons in nuclei,''
Nucl.\ Phys.\ B {\bf 484} (1997) 265.
%[arXiv:hep-ph/9608322].
%%CITATION = HEP-PH 9608322;%%
%
%\cite{Salgado:2003gb}
\bibitem{Salgado:2003gb}
C.~A.~Salgado and U.~A.~Wiedemann,
%``Calculating quenching weights,''
Phys.\ Rev.\ D {\bf 68} (2003) 014008.
%[arXiv:hep-ph/0302184].
%%CITATION = HEP-PH 0302184;%%
%
%\cite{Guo:2000nz}
\bibitem{Guo:2000nz}
X.~F.~Guo and X.~N.~Wang,
%``Multiple scattering, parton energy loss and modified fragmentation  functions
%in deeply inelastic e A scattering,''
Phys.\ Rev.\ Lett.\  {\bf 85} (2000) 3591.
% [arXiv:hep-ph/0005044].
%%CITATION = HEP-PH 0005044;%%
%
%\cite{Luo:1993ui}
\bibitem{Luo:1993ui}
  M.~Luo, J.~W.~Qiu and G.~Sterman,
  %``Twist four nuclear parton distributions from photoproduction,''
  Phys.\ Rev.\ D {\bf 49} (1994) 4493.
  %%CITATION = PHRVA,D49,4493;%%
%
%\cite{Adler:2003ii}
\bibitem{Adler:2003ii}
  S.~S.~Adler {\it et al.}  [PHENIX Collaboration],
  %``Absence of suppression in particle production at large transverse  momentum
  %in s(NN)**(1/2) = 200-GeV d + Au collisions,''
  Phys.\ Rev.\ Lett.\  {\bf 91} (2003) 072303.
%  [arXiv:nucl-ex/0306021].
  %%CITATION = NUCL-EX 0306021;%%

%\cite{Kniehl:2000fe}
\bibitem{Kniehl:2000fe}
  B.~A.~Kniehl, G.~Kramer and B.~P\"otter,
  %``Fragmentation functions for pions, kaons, and protons at  next-to-leading
  %order,''
  Nucl.\ Phys.\ B {\bf 582} (2000) 514.
%  [arXiv:hep-ph/0010289].
  %%CITATION = HEP-PH 0010289;%%
%
\end{thebibliography}
\end{document}